\documentstyle[prd,aps,tighten,preprint]{revtex}

\begin{document}
\author{Sheng Li\thanks{%
lisheng@itp.ac.cn}}
\address{CCAST World Lab, Academia Sinica, Beijing 100080, P. R. China}
\address{Institute of Theoretical Physics, Academia Sinica, P. O. Box 2735,
Beijing 100080, P. R. China}
\title{The Topological Structure of Nieh-Yan Form and Chiral Anomaly in Spaces with
Torsion}
\draft
\maketitle

\begin{abstract}
The topological structure of the Nieh-Yan form in 4-dimensional manifold is
given by making use of the decomposition of spin connection. The case of the
generalized Nieh-Yan form on $2^d$-dimensional manifold is discussed with an
example of 8-dimensional case studied in detail. The chiral anomaly with
nonvanishing torsion is studied also. The further contributions from
torsional part to chiral anomaly are found coming from the zeroes of some
fields under pure gauge condition.
\end{abstract}



\section{Introduction}

Torsion might be the most unusual field in physics. Though it has been under
investigation for more than two thirds of a century, there is no general
agreement on its mathematical formulation nor on its physical significance.
Vast funds of work on torsion had been done by many physicists (see \cite
{ein,sci,ham1,ram,zanelli,obukhov,obukhov2,mielke,bellisani,zhang,soo} for
examples) since it was proposed by E. Cartan\cite{cartan} in 1920s.

Consider a compact manifold $M$ with metric $g_{\mu \nu }$. There are two
dynamically independent one-forms: the connection $\omega ^{ab}=\omega _\mu
^{ab}dx^\mu $ and vielbein $e^a=e_\mu ^adx^\mu $. The curvature and torsion
2-forms are defined by 
\begin{eqnarray}
R^{ab} &=&d\omega ^{ab}-\omega ^{ac}\wedge \omega ^{cb} \\
T^a &=&de^a-\omega ^{ab}\wedge e^b.
\end{eqnarray}
In geometry, $\omega ^{ab}$ and $e^a$ reflect the affine and metric
properties of $M$. While in physics, curvature and torsion may be related to
energy momentum tensor and spin current respectively.

Nieh and Yan\cite{nieh-yan} firstly gave the 4-dimensional torsional
invariant 4-form as 
\begin{equation}
N=T^a\wedge T^a-R^{ab}\wedge e^a\wedge e^b.
\end{equation}
This is the only nontrivial locally exact 4-from which vanishes in the
absence of torsion and is clearly independent of the Pontryagin and Euler
densities. In any local patch where the vielbein is well defined, $N$ can be
written as 
\begin{equation}
N=d(e^a\wedge T^a)  \label{n-y}
\end{equation}
and therefore is locally exact. The 3-form $e^a\wedge T^a$ is a
Chern-Simons-like form that can be used as a Lagrangian for the dreibein in
three dimensions. The dual of this 3-form in four dimensions is also known
as the totally antisymmetric part of the torsion and is sometimes also
referred to as H-torsion, 
\begin{equation}
e^a\wedge T^a\wedge dx^\rho =\varepsilon ^{\mu \nu \lambda \rho }T_{\mu
\nu \lambda }d^4x.
\end{equation}
This component of the torsion tensor is the one that couples to the spin 1/2
fields\cite{obukhov2}.

Recently, there have been some discussions on the question of further
contributions to the Chiral anomaly in the presence of space-time with
torsion \cite{zanelli,obukhov,obukhov2,mielke,bellisani,zhang,soo}. It is
shown the Nieh-Yan form does contribute to the chiral anomaly for massive
field. This further anomaly term is associated with vacuum polarization
diagrams with two external axial torsion vertices, rather than with the
usual triangle diagrams\cite{soo}.

In this paper, by making use of the decomposition theory of $SO(N)$ spin
connection (see \cite{lisheng} for example) reviewed in section 2, we give
the topological structure of Nieh-Yan form in 4-dimensional manifold in
section 3. The Nieh-Yan number is found to be the sum of indices of some
field $\phi $ at its zeroes. The Hopf indices and Brouwer degrees of $\phi $
label the local properties of the Nieh-Yan form. We also present the
relationship between Nieh-Yan number and winding number. In section 4, a
general discussion on the cases of generalized Nieh-Yan form on $2^d$%
-dimensional manifold is given with a elaborate study on 8-dimensional case
as an example. By applying the results obtained in this paper and the
results obtained by Duan and Fu in ref. \cite{duanfu} to chiral anomaly, it
is found under pure gauge condition, contributions by chiral anomaly come
only from the zeroes of some field $\tilde{\phi}_{L(R)}$ and $\phi $ in
section 5. At last, we give a short conclusion in section 6. One should
notice that the main result of this paper is based on reduction to
contribution by singular points, i.e. those where ordinary formulas do not
hold.

\section{The decomposition theory of spin connection}

In this section, we give a short review of the decomposition theory of $%
SO(N) $ which is a useful tool in the discussion of the topological
structure of Nieh-Yan form in the later sections.

A smooth vector field $\varphi ^a$ $(a=1,2,\cdots ,N)$ can be found on the
base manifold ${\bf M}$ (a section of a vector bundle over ${\bf M}$). We
define a unit vector $n$ on ${\bf M}$ as 
\begin{equation}
n^a=\varphi ^a/||\varphi ||\qquad {}a=1,2,\cdots ,N  \label{f201}
\end{equation}
\[
||\varphi ||=\sqrt{\varphi ^a\varphi ^a}, 
\]
in which the superscript $``a"$ is the local orthonormal frame index. In
fact $n$ is identified as a section of the sphere bundle over ${\bf M}$ (or
a partial section of the vector bundle over ${\bf M}$). We see that the
zeroes of $\phi $ are just the singular points of $n$.

Let the $N$-dimensional Dirac matrix $\gamma _a$ ($a=1,2,\cdots ,N$) be the
basis of the Clifford algebra which satisfies 
\begin{equation}
\gamma _a\gamma _b+\gamma _b\gamma _a=2\delta _{ab}.  \label{f208}
\end{equation}
A unit vector field $n$ on ${\bf M}$ can be expressed as a vector of
Clifford Algebra 
\begin{equation}
n=n^a\gamma _a,  \label{f209}
\end{equation}
The spin connection $1$-form and curvature $2$-form are represented as
Clifford-algebra-valued differential forms respectively 
\begin{equation}
\omega =\frac 12\omega ^{ab}I_{ab}\qquad F={\frac 12}F^{ab}I_{ab},
\label{f210}
\end{equation}
in which $I_{ab}$ is the generator of the spin representation of the group $%
SO(N)$ 
\begin{equation}
I_{ab}={\frac 14}[\gamma _a,\gamma _b]={\frac 14}(\gamma _a\gamma _b-\gamma
_b\gamma _a).  \label{f211}
\end{equation}
The covariant derivative $1$-form of $n^a$ can be represented in terms of $n$
and $\omega $ as 
\begin{equation}
Dn=dn-[\omega ,n],  \label{f212}
\end{equation}
and curvature $2$-form is 
\begin{equation}
F=d\omega -\omega \wedge \omega .  \label{f213}
\end{equation}

Arbitrary $U\in Spin(N)$, which satisfies 
\begin{equation}
UU^{\dag }=U^{\dag }U=I,  \label{f216}
\end{equation}
is an even-versor\cite{hest}. The induced `spinorial' transformation by $U$
to the basis $\gamma _i$ of the Clifford algebra give $N$ orthonormal
vectors $u_i$\cite{Boer} via 
\begin{equation}
u_i:=U\gamma _iU^{\dag }=u_i^a\gamma _a,  \label{f217}
\end{equation}
where $u_i^a$ is the coefficient of $u_i$ in the representation of Clifford
algebra. From the relationship between $U$ and $u_i^a$, we see that $u_i$
has the same singular points with respect to different $``i"$. By (\ref{f217}%
), it is easy to verify that $u_i$ satisfy 
\begin{equation}
u_iu_j+u_ju_i=2\delta _{ij},\quad \quad \quad i,j=1,2,\cdots N.  \label{comm}
\end{equation}
From (\ref{f212}) we know that the covariant derivative 1-form of $u_i$ is 
\begin{equation}
Du_i=du_i-[\omega ,u_i].  \label{f222}
\end{equation}

There exists the following formula for a Clifford Algebra $r$-vector $A$\cite
{hest} 
\begin{equation}
u_iAu_i=(-1)^r(N-2r)A.  \label{comm1}
\end{equation}
For $\omega $ is a Clifford Algebra $2$-vector and using (\ref{comm1}), the
spin connection $\omega $ can be decomposed by $N$ orthonormal vectors $u_i$
as 
\begin{equation}
\omega ={\frac 14}(du_iu_i-Du_iu_i)  \label{f225}
\end{equation}
or 
\begin{equation}
\omega ^{ab}=du_i^au_i^b-Du_i^au_i^b.
\end{equation}
It can be proved that the general decomposition formula (\ref{f225}) has
global property and is independent of the choice of the local coordinates%
\cite{lisheng}.

By choosing the gauge condition 
\begin{equation}
Du_i=0,  \label{gaugecondition}
\end{equation}
we can define a generalized pseudo-flat spin connection as 
\begin{equation}
\omega _0=\frac 14du_iu_i.  \label{pseudo}
\end{equation}
Suppose there exist $l$ singular points $z_i$ ($i=1,2,\cdots ,l.$) in the
orthonormal vectors $u_j$. One can easily prove that at the normal points of 
$u_j$ 
\begin{equation}
F(\omega _0)=0\quad \quad \quad when\quad x\neq z_i.  \label{fis0}
\end{equation}
For the derivative of $u_i$ at the singular points $z_i$ is undefined, 
formula (\ref{fis0}) is invalid at $z_i$. Hence, the curvature under the
gauge condition (\ref{gaugecondition}) is a generalized function 
\begin{equation}
F\left\{ 
\begin{array}{c}
=0 \\ 
\neq 0
\end{array}
\quad \quad \quad when\quad 
\begin{array}{c}
x\neq z_i, \\ 
x=z_i.
\end{array}
\right. .  \label{f235}
\end{equation}
This is why we call $\omega _0$ the pseudo-flat spin connection. In fact,
the gauge condition (\ref{gaugecondition}) is the pure gauge condition
supported by the fact that one can always find a frame which is locally flat.

\section{Topological structure of Nieh-Yan form in 4-dimensional manifold}

One of the properties of topological invariant is that it is independent of
connection. Therefore, we choose the pseudo-flat connection 
\begin{equation}
\omega ^{ab}=du_i^au_i^b
\end{equation}
to make the calculation easier.\ Under this gauge condition, there must
exist singularity points on the manifold if the topology of this manifold is
nontrivial. In fact, what we have done is to choose a frame which is locally
flat. The choice of frame does not change the topology of the manifold. For
the singular property of the pseudo-flat connection, the contribution by the
zeros of some field at its singular points will give the topological
characteristic. It is a useful method to choose a locally flat frame when
dealing with the topological properties of manifolds, see for example\cite
{chen}.

Using the pseudo-flat spin connection, we can rewrite torsion as 
\begin{equation}
T^a=De^a=D(e_iu_i^a)=de_iu_i^a
\end{equation}
and Nieh-Yan form 
\begin{equation}
N=d(e^a\wedge T^a)=de_i\wedge de_i,
\end{equation}
where $e_i$ are the projection of vielbein onto the basis 
\begin{equation}
e_i=e^au_i^a.
\end{equation}

In quaternionic representation, a four vector $\vec{\phi}=(\phi _1,\phi
_2,\phi _3,\phi _4)$ is written as 
\begin{equation}
\phi =\phi _1i+\phi _2j+\phi _3k+\phi _4\quad \quad \phi ^{*}=-\phi _1i-\phi
_2j-\phi _3k+\phi _4
\end{equation}
and the vielbein projection 
\begin{equation}
e=e_1i+e_2j+e_3k+e_4,
\end{equation}
where $(1,i,j,k)$ is the basis of quaternion satisfying $i^2=-1$, $ij=k$,
etc. To give a frame that is flat locally, we can find some $\phi $, which
make the vielbein expressed in the form 
\begin{equation}
e=\frac l{||\phi ||^2}\phi d\phi ^{*},
\end{equation}
where the constant $l$ has dimension of length. The choice of local flat
metric is coincide with the pseudo-flat connection chosen, which makes
singular points exist for the non-trivial topological property of the
manifold..

It's well known that the quaternionic representation can be expressed in
terms of the Clifford algebra as 
\begin{equation}
\phi =\phi _is^i\quad \quad i=1,2,3,4,
\end{equation}
where 
\begin{equation}
s=(i\vec{\sigma},I),\quad \quad s^{\dagger }=(-i\vec{\sigma},I)
\end{equation}
and 
\begin{equation}
e=e_is^i.
\end{equation}
The vielbein projection can be rewritten as 
\begin{equation}
e=\frac l{||\phi ||^2}\phi d\phi ^{\dagger }.
\end{equation}
Then we get the Nieh-Yan form as 
\begin{equation}
N=de_i\wedge de_i=\frac 12Tr(de\wedge de).
\end{equation}
By making use of the relationship 
\begin{equation}
\varepsilon ^{ijkl}=\frac 12Tr(s^is^js^ks^l),
\end{equation}
the Nieh-Yan form can be expressed in terms of a unit vector $n_i=%
{\displaystyle {\phi _i \over ||\phi ||}}%
$ as 
\begin{eqnarray}
N &=&\frac{l^2}2Tr(dn\wedge dn\wedge dn\wedge dn)  \nonumber \\
&=&l^2\varepsilon ^{ijkl}dn_i\wedge dn_j\wedge dn_k\wedge dn_l.  \label{f323}
\end{eqnarray}

The derivative of $n_a$ can be deduced as 
\begin{equation}
dn_i=\frac{d\phi _i}{||\phi ||}-\phi _id({\frac 1{||\phi ||}}).
\end{equation}
Substituting it into (\ref{f323}), we have the expression of $N$ on $S({\bf M%
})$ 
\begin{equation}
N=l^2\varepsilon ^{ijkl}d(\frac{\phi _i}{||\phi ||^4}d\phi _j\wedge d\phi
_k\wedge d\phi _l).  \label{f3271}
\end{equation}
Using 
\begin{equation}
\frac{\phi _i}{||\phi ||^4}=-\frac 12\frac \partial {\partial \phi _i}(\frac %
1{||\phi ||^2}),
\end{equation}
equation (\ref{f3271}) is 
\begin{eqnarray}
N &=&-\frac{l^2}2\varepsilon ^{ijkl\ }\frac \partial {\partial \phi _m}\frac %
\partial {\partial \phi _i}(\frac 1{||\phi ||^2})  \nonumber \\
&&\times \frac{\partial \phi _m}{\partial {x}^\mu }\frac{\partial \phi _j}{%
\partial {x}^\nu }\frac{\partial \phi _k}{\partial {x}^\lambda }\frac{%
\partial \phi _l}{\partial {x}^\rho }\frac{\epsilon ^{\mu \nu \lambda \rho }%
}{\sqrt{g}}\sqrt{g}d^4x,
\end{eqnarray}
where $g=det(g_{\mu \nu })$, $g_{\mu \nu }$ is the metric tensor of ${\bf M}$%
. Define the Jacobian $D(\phi /x)$ as 
\begin{equation}
\varepsilon ^{ijkl}D(\phi /x)=\epsilon ^{\mu \nu \lambda \rho }\frac{%
\partial \phi _i}{\partial {x}^\mu }\frac{\partial \phi _j}{\partial {x}^\nu 
}\frac{\partial \phi _k}{\partial {x}^\lambda }\frac{\partial \phi _l}{%
\partial {x}^\rho }.
\end{equation}
Noticing 
\begin{equation}
\varepsilon ^{ijkl}\varepsilon ^{mjkl}=6\delta ^{im},
\end{equation}
we get 
\begin{equation}
N=-3l^2\frac{\partial ^2}{\partial \phi _i\partial \phi _i}(\frac 1{||\phi
||^2})D(\frac \phi x)d^4x.
\end{equation}
The general Green's-function formula$\cite{Gelf}$ in $\phi $ space is 
\begin{equation}
\frac{\partial ^2}{\partial \phi _i\partial \phi _i}(\frac 1{||\phi ||^2}%
)=-4\pi ^2\delta (\phi ),  \label{f553}
\end{equation}
We obtain the new formulation of Nieh-Yan form in terms of $\delta $
function $\delta (\phi )$ 
\begin{equation}
N=12\pi ^2l^2\delta (\phi )D(\phi /x)d^4x.  \label{f328}
\end{equation}

Suppose $\phi (x)$ has $l$ isolated zeroes on ${\bf M}$ and let the $i$th
zero be $z_i$, it is well known from the ordinary theory of the $\delta $%
-function \cite{Schw} that 
\begin{equation}
\delta (\phi )=\sum\limits_{i=1}^l\frac{\beta _i\delta (x-z_i)}{D(\phi
/x)|_{x=z_i}}.  \label{f330}
\end{equation}
Then one obtains 
\begin{equation}
\delta (\phi )D(\frac \phi x)=\sum\limits_{i=1}^l\beta _i\eta _i\delta
(x-z_i),  \label{f331}
\end{equation}
where $\beta _i$ is the positive integer (the Hopf index of the $i$th zero)
and $\eta _i$ the Brouwer degree\cite{Dubr,Miln1} 
\begin{equation}
\eta _i=sgnD(\phi /x)|_{x=z_i}=\pm 1.
\end{equation}
From the above deduction the following topological structure is obtained: 
\begin{equation}
n_{N-Y}=12\pi ^2l^2\delta (\phi )D(\frac \phi x)d^4x=12\pi
^2l^2\sum\limits_{i=1}^l\beta _i\eta _i\delta (x-z_i)d^4x,  \label{f332}
\end{equation}
which means that the local structure of $N$ is labeled by the Brouwer
degrees and Hopf indices, which are topological invariants. Therefore the
Nieh-Yan number $n_{N-Y}$ can be represented as 
\begin{equation}
n_{N-Y}=\int_{{\bf M}}N=12\pi ^2l^2\sum\limits_{i=1}^l\beta _i\eta _i.
\label{f333}
\end{equation}

On another hand, we can decompose ${\bf M}$ as 
\begin{equation}
{\bf M}=\sum_i{\bf M}_i,
\end{equation}
so that ${\bf M}_i$ includes only the $i$th singularity point $z_i$ of $n(x)$%
. Then we get 
\begin{eqnarray}
n_{N-Y} &=&\sum_i\int_{{\bf M}_i}l^2\varepsilon ^{ijkl}dn_i\wedge dn_j\wedge
dn_k\wedge dn_l  \nonumber \\
&=&\sum_i\oint_{\partial {\bf M}_i}l^2\varepsilon ^{ijkl}n_i\wedge
dn_j\wedge dn_k\wedge dn_l,  \label{winding}
\end{eqnarray}
where $\partial {\bf M}_i$ is the boundary of ${\bf M}_i$. Equation (\ref
{winding}) is another definition of the winding number $W(\phi ,z_i)$ of the
surface $\partial {\bf M}_i$ and the mapping $\phi (x)$\cite{victor} 
\begin{equation}
W(\phi ,z_i)=\frac 1{12\pi ^2}\oint_{\partial {\bf M}_i}\varepsilon
^{ijkl}n_i\wedge dn_j\wedge dn_k\wedge dn_l=\beta _i\eta _i.
\end{equation}
Then the Nieh-Yan number $n_{N-Y}$ can further be expressed in terms of
winding numbers 
\begin{equation}
n_{N-Y}=12\pi ^2l^2\sum_{i=1}^lW(\phi ,z_i).
\end{equation}
The sum of the winding numbers can be interpreted or, indeed, defined as the
degree of the mapping $\phi (x)$ onto ${\bf M}$. By (\ref{f328}) and (\ref
{winding}), we have 
\begin{eqnarray}
\sum_{i=1}^lW(\phi ,z_i) &=&\int_{{\bf M}}\delta (\phi )D(\frac \phi x)d^4x 
\nonumber \\
&=&\deg \phi \int_{\phi ({\bf M})}\delta (\phi )d^4\phi  \nonumber \\
&=&\deg \phi .
\end{eqnarray}
Therefore, we get the Nieh-Yan number at last 
\begin{equation}
n_{N-Y}=12\pi ^2l^2\sum_{i=1}^lW(\phi ,z_i)=12\pi ^2l^2\deg \phi .
\label{nylast}
\end{equation}

From (\ref{f201}), we know that the zeroes of $\phi $ are just the
singularities of $n$. Here (\ref{f333}) says that the sum of the indices of
the singular points of $n,$ or of the zeroes of $\phi $, is Nieh-Yan number.
One should notice that the vector field $\phi $ is not a section of the
tangent bundle of ${\bf M}$, i.e. $\phi $ is not a Riemannian vector field.
Therefore formula (\ref{nylast}) does not concern with the Poincar\'{e}-Hopf
theorem which connects the Euler characteristics with the zeros of a
Riemannian vector field.

\section{Higher dimensional case}

In $2^d$-dimensional case, we can get similar results. Under the pure gauge
condition, the nonvanishing term of the generalized Nieh-Yan form is 
\begin{equation}
(de_i\wedge de_i)^{2^{d-2}}.
\end{equation}
The computation above is also valid in this case. The topological structure
of generalized Nieh-Yan form is constitute by delta function of some $\phi $%
. The Hopf indices and Brouwer degree labeled the local structure of
Nieh-Yan form. The Nieh-Yan number is some constant number times of the
degree or winding numbers of $\phi $. 
\begin{equation}
N\sim \delta (\phi )D(\frac \phi x)d^{2^d}x=\sum_i\beta _i\eta _i\delta
(x-z_i)d^{2^d}x.
\end{equation}
In the following, we take only 8-dimensional case as an example.

In 8-dimensional compact manifold, under the gauge condition (\ref
{gaugecondition}), the generalized Nieh-Yan form is expressed simply as 
\begin{equation}
N=d(e_i\wedge de_i\wedge de_j\wedge de_j).  \label{8dny}
\end{equation}
And the vielbein is expressed locally in terms of the octonions under
Clifford algebra representation 
\begin{equation}
e=e_is^i=\frac l{||\phi ||^2}\phi d\phi ^{\dagger },  \label{8-de}
\end{equation}
where 
\[
\phi =\phi _is^i\quad \quad i=1,2,...,8. 
\]
in which $s^i$\ is the basis of octonions. Substituting (\ref{8-de}) into (%
\ref{8dny}), we get the Nieh-Yan form of 8-d in terms of unit vector $n_i={{%
\frac{\phi _i}{||\phi ||}}}$ as 
\begin{equation}
N=\frac{l^4}5\varepsilon ^{i_1i_2...i_8}dn_{i_1}\wedge dn_{i_2}\wedge
...\wedge dn_{i_8}.
\end{equation}
Analogously to 4-dimensional case, using the general Green's-function formula%
\cite{Gelf} in 8-dimensional $\phi $ space, the generalized Nieh-Yan form is
expressed as 
\begin{eqnarray}
N &=&1008l^4A(S^7)\delta (\phi )D(\frac \phi x)d^8x  \nonumber \\
&=&336\pi ^4l^4\delta (\phi )D(\frac \phi x)d^8x,
\end{eqnarray}
which can be rewritten in terms of Hopf indices and Brouwer degree as 
\begin{equation}
N=336\pi ^4l^4\sum_i\beta _i\eta _i\delta (x-z_i)d^8x.
\end{equation}
The Nieh-Yan number in 8-dimensional is 
\begin{equation}
n_{N-Y}=336\pi ^4l^4\deg \phi =336\pi ^4l^4\sum_iW(\phi ,z_i).
\end{equation}

Similar results can be obtained for higher $2^d$-dimensional case.

\section{Chiral anomaly on spaces with torsion}

Consider a massive Dirac spinor on a curved background with torsion. The
action is 
\begin{equation}
S=\int \frac 12(d^4xe\bar{\psi}\ \not{\nabla}\psi +h.c.)+m\bar{\psi}\psi ,
\end{equation}
where the Dirac operator is given by 
\begin{equation}
\not{\nabla}=e^{\mu a}\gamma _aD_\mu .
\end{equation}
This action is invariant under rigid chiral transformations 
\begin{equation}
\psi ^{\prime }\rightarrow e^{i\varepsilon \gamma _5}\psi ,
\end{equation}
where $\varepsilon $ is a real constant parameter. This symmetry leads to
the classical conservation law 
\begin{equation}
\partial _\mu J_5^\mu =0,
\end{equation}
in which 
\begin{equation}
J_5^\mu =ee^{\mu a}\bar{\psi}\gamma _a\gamma _5\psi .
\end{equation}

The chiral anomaly when torsion is present is given by\cite
{zanelli,soo,obukhov,mielke} 
\begin{equation}
\partial _\mu <J_5^\mu >=A(x),
\end{equation}
with 
\begin{equation}
A(x)=\frac 1{8\pi ^2}*[R^{ab}\wedge R^{ab}+\frac 2{l^2}(T^a\wedge
T^a-R^{ab}\wedge e^a\wedge e^b)].
\end{equation}
The constant $l$ is called the radius of the universe and is related to the
cosmological constant ($|\Lambda |=l^{-2}$).

Using the results obtained in section 3, we can rewritten the chiral anomaly
as 
\begin{equation}
\partial _\mu <J_5^\mu >=\frac 1{8\pi ^2}*[R^{ab}\wedge R^{ab}]+3\delta
(\phi )D(\frac \phi x).
\end{equation}
From the relationship 
\begin{equation}
so(4)=su(2)_L\times su(2)_R,
\end{equation}
the Pontryagin class of $SO(4)$ group can be expressed as the sum of the
second Chern classes of the left and right $SU(2)_{L(R)}$ subgroup 
\begin{equation}
Tr(R\wedge R)=Tr(R_{SU(2)_L}\wedge R_{SU(2)_L})+Tr(R_{SU(2)_R}\wedge
R_{SU(2)_R}).
\end{equation}
By making use of the result of Duan and Fu\cite{duanfu} 
\begin{equation}
\frac 1{8\pi ^2}Tr(R_{SU(2)}\wedge R_{SU(2)})=\delta (\tilde{\phi})D(\frac{%
\tilde{\phi}}x)d^4x,
\end{equation}
where $\grave{\phi}$ belongs to $Spin(3)$ corresponding to the group $SU(2)$%
. In Duan and Fu's paper, the pure gauge condition is used also to get the
topological structure of the second Chern class for $SU(2)$ group. Now we
get the chiral anomaly 
\begin{equation}
\partial _\mu <J_5^\mu >=\delta (\tilde{\phi}_L)D(\frac{\tilde{\phi}_L}x%
)+\delta (\tilde{\phi}_R)D(\frac{\tilde{\phi}_R}x)+3\delta (\phi )D(\frac %
\phi x),
\end{equation}
in which $\tilde{\phi}_L$ and $\tilde{\phi}_R$ are some $Spin(3)_{L(R)}$
elements corresponding to the subgroups $SU(2)_{L(R)}$. Furthermore, by
making use of the structure of delta function, the anomaly can be formulated
more explicitly 
\begin{eqnarray}
\partial _\mu &<&J_5^\mu >=\sum_i\delta (x-z_{Li})\beta _{Li}\eta
_{Li}+\sum_i\delta (x-z_{Ri})\beta _{Ri}\eta _{Ri}+3\sum_i\delta
(x-z_i)\beta _i\eta _i  \nonumber \\
&=&\sum_i\delta (x-z_{Li})W_{Li}+\sum_i\delta (x-z_{Ri})W_{Ri}+3\sum_i\delta
(x-z_i)W_i.  \label{calast}
\end{eqnarray}
where $\beta _{L(R)i}$ and $\eta _{L(R)i}$ are the Hopf indices and Brouwer
degrees respectively corresponding to the zeroes of $\tilde{\phi}_{L(R)}$,
and $W_{L(R)i}$ is the winding number of $\tilde{\phi}_{L(R)}$ at its $i$th
zeroes. From (\ref{calast}), we see, under pure gauge condition, the chiral
anomaly comes only from the zeroes of the fields $\tilde{\phi}_{LR}$ and $%
\phi $, and their winding numbers account the quantity of chiral anomaly.

\section{Conclusion}

In this paper, we have discussed the Nieh-Yan form by making use of the
decomposition theory of spin connection. The Nieh-Yan form in 4-dimensional
manifold is proved to take the delta function form under local flat gauge
condition. The local topological structure of Nieh-Yan form is labeled by
Hopf indices and Brouwer degrees of field $\phi $, which is used to
expressed the vielbein under local flat (or pure) gauge condition. The
Nieh-Yan number is proved to be a constant number times the degree or
winding number of $\phi $.

A general discussion of the generalized Nieh-Yan form on $2^d$-dimensional
manifold is presented. From an example of 8-dimensional case, we find the
topological structure of the generalized Nieh-Yan form is similar to the
4-dimensional case in terms of field $\phi $ under pure gauge condition. It
is noticeable that the Clifford algebra can make the calculus more easier in
these cases.

The topological anomaly with nonvanishing torsion is formulated explicitly
under pure gauge condition. There are two kinds of contribution in the
anomaly which comes from the topology of the manifold: the Pontryagin class
of $SO(4)$ and the Nieh-Yan form. We have proved each of them to be a sum of
delta function of some fields $\tilde{\phi}_{L(R)}$ and $\phi $ under pure
gauge condition. It means the contribution comes only from the zeroes of $%
\tilde{\phi}_{L(R)}$ and $\phi $. The degrees or winding numbers of $\tilde{%
\phi}_{L(R)}$ or $\phi $ give the quantity of contributions.

\end{document}